\documentstyle[amssymb,11pt]{article}

\newlength{\minitwocolumn}
\font\teneufm=eufm10
\font\seveneufm=eufm7
\font\fiveeufm=eufm5
\newfam\eufmfam
\textfont\eufmfam=\teneufm
\scriptfont\eufmfam=\seveneufm
\scriptscriptfont\eufmfam=\fiveeufm

\let\goth\frak
\makeatletter
\@addtoreset{equation}{section}
\makeatother

\newtheorem{thm}{Theorem}[section]
\newtheorem{prop}[thm]{Proposition}

\newtheorem{df}{Definition}[section]

\title{\bf
\huge{\bf
Wakimoto realization of\\ 
the elliptic algebra $U_{q,p}(\widehat{sl_N})$}
}
\begin{document}
\maketitle
\begin{center}
{Takeo KOJIMA}
\\~\\
{\it
Department of Mathematics,
College of Science and Technology,
Nihon University,\\
Surugadai, Chiyoda-ku, Tokyo 101-0062, 
JAPAN}
\end{center}
~\\
\begin{abstract}
We construct a free field realization of
the elliptic quantum algebra
$U_{q,p}(\widehat{sl_N})$ for arbitrary level $k \neq 0,-N$.
We study Drinfeld current
and the screening current associated with $U_{q,p}(\widehat{sl_N})$
for arbitrary level $k$.
In the limit $p \to 0$ this realization becomes $q$-Wakimoto
realization for $U_q(\widehat{sl_N})$.
\end{abstract}

~\\

\section{Introduction}

The elliptic quantum group has been proposed in papers 
\cite{FIJKMY,
Felder, Fronsdal, EF, JKOS1}.
There are two types of elliptic quantum groups,
the vertex type ${\cal A}_{q,p}(\widehat{sl_N})$ and the face type
${\cal B}_{q,\lambda}({\goth g})$, where ${\goth g}$ is a Kac-Moody algebra
associated with a symmetrizable Cartan matrix.
Not only the quantum group
but also
the elliptic quantum groups have the structure of quasi-triangular quasi-Hopf algebras
introduced by V.Drinfeld \cite{Drinfeld}.
H.Konno \cite{Konno} introduced the elliptic quantum algebra $U_{q,p}(\widehat{sl_2})$ 
as an algebra of the elliptic analogue of Drinfeld current
in the context of the fusion SOS model \cite{DJKMO}.
M.Jimbo, H.Konno, S.Odake, J.Shiraishi
\cite{JKOS2} continued to study the elliptic quantum algebra $U_{q,p}(\widehat{sl_2})$.
They identified $U_{q,p}(\widehat{sl_2})$ with
the tensor product of ${\cal B}_{q,\lambda}(\widehat{sl_2})$ and a Heisenberg algebra
${\cal H}$.
The elliptic quantum group ${\cal B}_{q,\lambda}(\widehat{sl_2})$ is a quasi-Hopf algebra
while the elliptic algebra $U_{q,p}(\widehat{sl_2})$ is not.
The intertwining relation of the vertex operator of
${\cal B}_{q,\lambda}(\widehat{sl_2})$ is based on the quasi-Hopf structure of
${\cal B}_{q,\lambda}(\widehat{sl_2})$.
By the above isomorphism $U_{q,p}(\widehat{sl_2})\simeq {\cal B}_{q,\lambda}
(\widehat{sl_2}) \otimes {\cal H}$,
we can understand "intertwining relation" of
the vertex operator for the elliptic algebra $U_{q,p}(\widehat{sl_2})$.
Going along the isomorphism
$U_{q,p}({\goth g})\simeq {\cal B}_{q,\lambda}
({\goth g}) \otimes {\cal H}$,
 the elliptic analogue of
Drinfeld current of $U_{q,p}(\widehat{sl_2})$ is extended to those of
$U_{q,p}({\goth g})$ for non-twisted affine Lie algebra ${\goth g}$ \cite{JKOS2, KK}.
In this paper we are interested in higher-rank generalization
of level $k$ free field realization of
the elliptic quantum algebra.
For the elliptic algebra
$U_{q,p}(\widehat{sl_2})$,
there exist two kind of free field realizations
for arbitrary level $k$,
the one is parafermion realization \cite{Konno, JKOS2}, the other is
Wakimoto realization \cite{CD}.
In this paper we are interested in the higher-rank generalization of
Wakimoto realization of $U_{q,p}(\widehat{sl_2})$.
We construct level $k$ free field realization of Drinfeld current
associated with the elliptic algebra $U_{q,p}(\widehat{sl_N})$.
This gives the higher-rank generalization of
the author's previous work on $U_{q,p}(\widehat{sl_3})$ \cite{Kojima}.
It is supposed that this free field realization can be applied for
construction of the level $k$ integrals of motion
for the elliptic algebra $U_{q,p}(\widehat{sl_N})$.
For this purpose, see references \cite{KS1, KS2, KS3}.

The organization of this paper is as follows.
In section 2 we set the notation and introduce bosons.
In section 3 we review the level $k$ free field realization of the 
quantum group $U_q(\widehat{sl_N})$ \cite{AOS, AOS2}.
In section 4 we give the free field realization of the dressing operator 
$U^i(z), U^{*i}(z)$, which cause the elliptic deformation of Drinfeld current.
We study the screening current, too.
In appendix A, we explain a systematic way of construction
for a free field realization of
the dressing operators $U^i(z)$ and $U^{*i}(z)$.
In appendix B, we summarize the normal ordering of the basic operators.

~

After finishing this work, I noticed 
a paper on $U_{q,p}(\widehat{sl_N})$
by W.Chang and X.Ding \cite{CD}
[math.QA:0812.1147],
which seemed to be submitted to arXiv. a day after my submitting
$U_{q,p}(\widehat{sl_3})$ paper
\cite{Kojima}
[nlin.SI:0812.0890,
proceedings of the 27-th International Colloquium, Armenia, August 2008].

\section{Bosons}
The purpose of this section is to set up the basic notation and
to introduce the boson.
In this paper we fix three parameters $q,k,r \in {\mathbb C}$.
Let us set $r^*=r-k$.
We assume $k \neq 0, -N$ and ${\rm Re}(r)>0$, ${\rm Re}(r^*)>0$.
We assume $q$ is generic with $|q|<1, q \neq 0$.
Let us set a pair of parameters $p$ and $p^*$ by
\begin{eqnarray}
p=q^{2r},~~p^*=q^{2r^*}.\nonumber
\end{eqnarray}
We use the standard symbol of $q$-integer $[n]$ by
\begin{eqnarray}
[n]=\frac{q^n-q^{-n}}{q-q^{-1}}.\nonumber
\end{eqnarray}
Following \cite{AOS, AOS2} we introduce free bosons 
$a_n^i,(1\leqq i \leqq N-1; n \in {\mathbb Z}_{\neq 0})$,
$b_n^{i,j}, (1\leqq i<j \leqq N; n \in {\mathbb Z}_{\neq 0})$,
$c_n^{i,j}, (1\leqq i<j \leqq N; n \in {\mathbb Z}_{\neq 0})$,
and the zero-mode operators
$a^i,(1\leqq i \leqq N-1)$,
$b^{i,j},(1\leqq i<j \leqq N)$,
$c^{i,j},(1\leqq i<j \leqq N)$.
\begin{eqnarray}
~[a_n^i,a_m^j]&=&\frac{[(k+N)n][A_{i,j}n]}{n}\delta_{n+m,0},~~
[p_a^i,q_a^j]=(k+N) A_{i,j},
\\
~[b_n^{i,j},b_m^{k,l}]&=&-\frac{[n]^2}{n} 
\delta_{i,k}
\delta_{j,l}
\delta_{n+m,0},~~
[p_b^{i,j},q_b^{k,l}]=-\delta_{i,k}\delta_{k,l},
\\
~[c_n^{i,j},c_m^{k,l}]&=&
\frac{[n]^2}{n} \delta_{i,k}\delta_{j,l} \delta_{n+m,0},~~
[p_c^{i,j},q_c^{k,l}]=\delta_{i,k}\delta_{j,l}.
\end{eqnarray}
Here the matrix $(A_{i,j})_{1\leqq i,j \leqq N-1}$
represents the Cartan matrix of classical $sl_N$.
For parameters 
$a_i \in {\mathbb R},(1\leqq i \leqq N-1)$, 
$b_{i,j} \in {\mathbb R},(1\leqq i<j \leqq N)$
$c_{i,j},\in {\mathbb R}, (1\leqq i<j \leqq N)$,
we set
the vacuum vector $|a, b, c\rangle$ of the Fock space 
${\cal F}_{a,b,c}$ as following.
\begin{eqnarray}
&&a_n^i|a,b,c\rangle=b_n^{j,k}|a,b,c\rangle
=c_n^{j,k}|a,b,c\rangle=0,~~(n>0; 1\leqq i \leqq N-1; 
1\leqq j<k \leqq N),\nonumber
\end{eqnarray}
\begin{eqnarray}
p_a^i|a,b,c\rangle=a_i|a,b,c\rangle,~
p_b^{j,k}|a,b,c\rangle=b_{j,k}|a,b,c\rangle,~
p_c^{j,k}|a,b,c\rangle=c_{j,k}|a,b,c\rangle,\nonumber
\\
~~(1\leqq i \leqq N-1; 1\leqq j<k \leqq N).\nonumber
\end{eqnarray}
The Fock space
${\cal F}_{a,b,c}$ 
is generated by bosons $a_{-n}^i,b_{-n}^{j,k},c_{-n}^{j,k}$ 
for $n \in {\mathbb N}_{\neq 0}$.
The dual Fock space
${\cal F}_{a,b,c}^*$ is defined as the same manner. 
In this paper we construct the elliptic analogue of Drinfeld current
for $U_{q,p}(\widehat{sl_N})$ by these bosons $a_n^i, b_n^{j,k}, c_n^{j,k}$
acting on the Fock space.

Let us set the elliptic theta function $\Theta_p(z)$ by
\begin{eqnarray}
\Theta_p(z)=(z;p)_\infty (p/z;p)_\infty (p;p)_\infty,~~~
(z;p)_\infty=\prod_{n=0}^\infty (1-p^nz).\nonumber
\end{eqnarray}
It is convenient to work with the additive notation.
We use the parametrization
\begin{eqnarray}
q&=&e^{-\pi \sqrt{-1}/r \tau},\nonumber\\
p&=&e^{-2\pi \sqrt{-1}/\tau},~~~
p^*=e^{-2\pi \sqrt{-1}/\tau^*},~~(r\tau=r^*\tau^*),\nonumber\\
z&=&q^{2u}.\nonumber
\end{eqnarray}
Let us set Jacobi elliptic theta function $[u]_r$ by
\begin{eqnarray}
~[u]_r&=&q^{\frac{u^2}{r}-u}\frac{\Theta_{q^{2r}}(z)}{(q^{2r};q^{2r})_\infty^3}.\nonumber
\end{eqnarray}
The function $[u]_r$ has a zero at $u=0$, enjoys the quasi-periodicity property
\begin{eqnarray}
~[u+r]_r=-[u]_r,~~~~[u+r\tau]_r=-e^{-\pi \sqrt{-1} \tau-
\frac{2\pi \sqrt{-1} u}{r}}[u]_r.\nonumber
\end{eqnarray}
Let us set the $q$-difference $(_{\alpha}\partial_z f)(z)$ by
\begin{eqnarray}
(_\alpha \partial_z f)(z)=\frac{f(q^\alpha z)-f(q^{-\alpha}z)}{(q-q^{-1})z}.\nonumber 
\end{eqnarray}
Let us set the delta-function $\delta(z)$ as formal power series.
\begin{eqnarray}
\delta(z)=\sum_{n \in {\mathbb Z}} z^n.\nonumber
\end{eqnarray}

\section{Free field realization of $U_q(\widehat{sl_N})$}

The purpose of this section is to give a review on
the free field realization of 
the quantum affine algebra $U_{q}(\widehat{sl_N})$
\cite{AOS2}, which is a basis of those of the elliptic algebra
$U_{q,p}(\widehat{sl_N})$.

\subsection{Drinfeld current}

Let us set the bosonic operators $a_\pm^i(z), a^i(z), (1\leqq i \leqq N-1)$, 
$b_\pm^{i,j}(z), b^{i,j}(z), c^{i.j}(z), (1\leqq i<j \leqq N)$ by
\begin{eqnarray}
a_\pm^i(z)&=&\pm(q-q^{-1})\sum_{n>0}a_{\pm n}^i z^{\mp n} \pm p_a^i {\rm log}q,
\\
b_\pm^{i,j}(z)&=&
\pm(q-q^{-1})\sum_{n>0}b_{\pm n}^{i,j} 
z^{\mp n} \pm p_b^{i,j} {\rm log}q,\\
a^i(z)&=&-\sum_{n \neq 0}\frac{a_n^i}{[(k+N)n]}q^{-\frac{k+N}{2}|n|}z^{-n}+
\frac{1}{k+N}(q_a^i+p_a^i {\rm log}z),\\
b^{i,j}(z)&=&-\sum_{n \neq 0}\frac{b_n^{i,j}}{[n]}z^{-n}+
q_b^{i,j}+p_b^{i,j}{\rm log}z,
\\
c^{i,j}(z)&=&-\sum_{n \neq 0}\frac{c_n^{i,j}}{[n]}z^{-n}+
q_c^{i,j}+p_c^{i,j}{\rm log}z,
\end{eqnarray}
Let us set the auxiliary operators $\gamma^{i,j}(z),\beta_1^{i,j}(z),
\beta_2^{i,j}(z), \beta_3^{i,j}(z), \beta_4^{i,j}(z)$,
$(1\leqq i<j \leqq N)$ by
\begin{eqnarray}
\gamma^{i,j}(z)&=&-\sum_{n \neq 0}\frac{(b+c)_n^{i,j}}{[n]}z^{-n}
+(q_b^{i,j}+q_c^{i,j})+(p_b^{i,j}+p_c^{i,j}){\rm log}(-z),\\
\beta_1^{i,j}(z)&=&b_+^{i,j}(z)-(b^{i,j}+c^{i,j})(qz),
~\beta_2^{i,j}(z)=b_-^{i,j}(z)-(b^{i,j}+c^{i,j})(q^{-1}z),
\\
\beta_3^{i,j}(z)&=&b_+^{i,j}(z)+(b^{i,j}+c^{i,j})(q^{-1}z),~
\beta_4^{i,j}(z)=b_-^{i,j}(z)+(b^{i,j}+c^{i,j})(qz).
\end{eqnarray}
We give a free field realization of Drinfeld current
for $U_q(\widehat{sl_N})$.

\begin{df}~~Let us set the bosonic operators
$E^{\pm,i}(z), (1\leqq i \leqq N-1)$ by
\begin{eqnarray}
E^{+,i}(z)=\frac{-1}{(q-q^{-1})z}\sum_{j=1}^i E^{+,i}_{j}(z),\\
E^{-,i}(z)=\frac{-1}{(q-q^{-1})z}\sum_{j=1}^{N-1} E^{-,i}_{j}(z),
\end{eqnarray}
where we have set
\begin{eqnarray}
E^{+,i}_j(z)=:e^{\gamma^{j,i}(q^{j-1}z)}(
e^{\beta_1^{j,i+1}(q^{j-1}z)}-e^{\beta_2^{j,i+1}(q^{j-1}z)})
e^{\sum_{l=1}^{j-1}(b_+^{l,i+1}(q^{l-1}z)-b_+^{l,i}(q^lz))}:,
\end{eqnarray}
\begin{eqnarray}
E^{-,i}_j(z)&=&:e^{\gamma^{j,i+1}(q^{-(k+j)}z)}
(e^{-\beta_4^{j,i}(q^{-(k+j)}z)}-e^{-\beta_3^{j,i}(q^{-(k+j)}z)})\nonumber\\
&\times& e^{\sum_{l=j+1}^i 
(b_-^{l,i+1}(q^{-(k+l-1)}z)-b_-^{l,i}(q^{-(k+l)}z))+a_-^i(q^{-\frac{k+N}{2}}z)
+\sum_{l=i+1}^N(b_-^{i,l}(q^{-(k+l)}z)-b_-^{i+1,l}(q^{-(k+l-1)}z))}:,
\nonumber\\
&&~~~~~~~~~~~~~~~~~~~{\rm for}~~~1\leqq j \leqq i-1,\\
E^{-,i}_i(z)&=&:e^{\gamma^{i,i+1}(q^{-(k+i)}z)+a_-^i(q^{-\frac{k+N}{2}}z)+
\sum_{l=i+1}^N (b_-^{i,l}(q^{-(k+l)}z)-b_-^{i+1,l}(q^{-(k+l-1)}z))}:\nonumber\\
&-&:e^{\gamma^{i,i+1}(q^{k+i}z)+a_+^i(q^{\frac{k+N}{2}}z)+\sum_{l=i+1}^N
(b_+^{i,l}(q^{k+l}z)-b_+^{i+1,l}(q^{k+l-1}z))}:,\\
E^{-,i}_j(z)&=&:e^{\gamma^{i,j+1}(q^{k+j}z)}
(e^{\beta_2^{i+1,j+1}(q^{k+j}z)}-e^{\beta_1^{i+1,j+1}(q^{k+j}z)})
e^{a_+^i(q^{\frac{k+N}{2}}z)+\sum_{l=j+1}^N (
b_+^{i,l}(q^{k+l}z)-b_+^{i+1,l}(q^{k+l-1}z))}:,\nonumber\\
&&~~~~~~~~~~~~~~~~~~~{\rm for}~~~i+1 \leqq j \leqq N-1.
\end{eqnarray}
Let us set the bosonic operators $\psi_i^\pm(z), (1\leqq i \leqq N-1)$ by
\begin{eqnarray}
\psi_\pm^i(q^{\pm \frac{k}{2}}z)=:e^{\sum_{j=1}^i (b_\pm^{j,i+1}
(q^{\pm (k+j-1)}z)-b_\pm^{j,i}(q^{\pm (k+j)}z))+a_\pm^i(q^{\pm 
\frac{k+N}{2}}z)+\sum_{j=i+1}^N 
(b_\pm^{i,j}(q^{\pm(k+j)}z)-b_\pm^{i+1,j}(q^{\pm (k+j-1)}z))}:.
\end{eqnarray}
Let us set
\begin{eqnarray}
h_i=\sum_{j=1}^i (p_b^{j,i+1}-p_b^{j,i})+p_a^i+\sum_{j=i+1}^N
(p_b^{i,j}-p_b^{i+1,j}).
\end{eqnarray}
\end{df}
Here the symbol $:{\cal O}:$ represents the normal ordering
of ${\cal O}$.
For example we have
\begin{eqnarray}
:b_k^{i,j} b_l^{i,j}:=\left\{
\begin{array}{cc}
b_k^{i,j} b_l^{i,j},& k<0\\
b_l^{i,j} b_k^{i,j},& k>0.
\end{array}\right.~~~
:p_b^{i,j} q_b^{i,j}:=:q_b^{i,j} p_b^{i,j}:=q_b^{i,j} p_b^{i,j}.\nonumber
\end{eqnarray}

\begin{thm}~~The operators $E^{\pm,i}(z), \psi_\pm^i(z), h_i$,
$(1\leqq i \leqq N-1)$
give a free field realization of $U_{q}(\widehat{sl_N})$ for arbitrary
level $k \neq 0, -N$. In other words, they satisfy
the following commutation relations.
\begin{eqnarray}
&&~[h_i,E^{\pm,j}(z)]=\pm A_{i,j} E^{\pm,j}(z),
\label{eqn:q1}
\\
&&(z_1-q^{\pm A_{i,j}}z_2)E^{\pm,i}(z_1)E^{\pm,j}(z_2)
=(q^{\pm A_{i,j}}z_1-z_2)
E^{\pm,j}(z_2)E^{\pm,i}(z_1),\label{eqn:q2}
\\
&&~[\psi_\pm^i(z_1),\psi_\pm^j(z_2)]=0,
\label{eqn:q3}
\\
&&(z_1-q^{A_{i,j}-k}z_2)(z_1-q^{-A_{i,j}+k}z_2)\psi_+^i(z_1)\psi_-^j(z_2)
\nonumber\\
&&=(z_1-q^{A_{i,j}+k}z_2)(z_1-q^{-A_{i,j}-k}z_2)\psi_-^j(z_2)\psi_+^i(z_1),
\label{eqn:q4}
\\
&&(z_1-q^{\pm (A_{i,j}-\frac{k}{2})}z_2)\psi_+^i(z_1)E^{\pm,j}(z_2)=
(q^{\pm A_{i,j}}z_1-q^{\mp \frac{k}{2}}z_2)E^{\pm,j}(z_2)\psi_+^i(z_1),
\label{eqn:q5}
\\
&&(z_1-q^{\pm (A_{i,j}-\frac{k}{2})}z_2)E^{\pm,i}(z_1)\psi_-^j(z_2)=
(q^{\pm A_{i,j}}z_1-q^{\mp \frac{k}{2}}z_2)\psi_-^j(z_2)E^{\pm,i}(z_1),
\label{eqn:q6}
\end{eqnarray}
\begin{eqnarray}
&&\left\{
E^{\pm,i}(z_1)E^{\pm,i}(z_2)E^{\pm,j}(z_3)-(q+q^{-1})
E^{\pm,i}(z_1)E^{\pm,j}(z_3)E^{\pm,i}(z_2)+
E^{\pm,i}(z_3)E^{\pm,i}(z_1)E^{\pm,j}(z_2)
\right\}\nonumber\\
&&+\left\{ z_1 \leftrightarrow z_2 \right\}=0,~~{\rm for}~~A_{i,j}=-1,
\label{eqn:q7}
\end{eqnarray}
\begin{eqnarray}
&&[E^{+,i}(z_1),E^{-,j}(z_2)]=\frac{\delta_{i,j}}{(q-q^{-1})z_1 z_2}
\left(\delta\left(q^{-k}\frac{z_1}{z_2}\right)\psi_+^i(q^{-\frac{k}{2}}z_1)
-\delta\left(q^k\frac{z_1}{z_2}\right)
\psi_-^i(q^{-\frac{k}{2}}z_2)\right).\nonumber\\
\label{eqn:q8}
\end{eqnarray}
\end{thm}
When we take the limit $q \to 1$, we recover Wakimoto realization for
$\widehat{sl_N}$ \cite{FF}.

\subsection{Screening current}

Following \cite{AOS2}, we define
the screening current $S^i(z)$, which commute with $U_q(\widehat{sl_N})$. 

\begin{df}~~Let us introduce the bosonic operator $S^i(z), (1\leqq i \leqq N-1)$
ny
\begin{eqnarray}
S^i(z)&=&\frac{-1}{(q-q^{-1})z}:e^{-a^i(z)}\widetilde{S}^i(z):,
\end{eqnarray}
where we have set
\begin{eqnarray}
\widetilde{S}^i(z)&=& \sum_{j=i+1}^N
:e^{\gamma^{i+1,j}(q^{N-j}z)}(
e^{-\beta_4^{i,j}(q^{N-j}z)}
-
e^{-\beta_3^{i,j}(q^{N-j}z)}
)e^{\sum_{l=j+1}^N
(b_-^{i+1,l}(q^{N-l+1}z)-b_-^{i,l}(q^{N-l}z))
}:.\nonumber
\end{eqnarray}
\end{df}

\begin{prop}~~
The bosonic operator $S^i(z), E^{\pm,i}(z)$ satisfy
the following commutation relations.
\begin{eqnarray}
&&\left[u_1-u_2-\frac{A_{i,j}}{2}\right]_{k+N} 
S^i(z_1)S^j(z_2)= 
\left[u_1-u_2+\frac{A_{i,j}}{2}\right]_{k+N} S^j(z_2)S^i(z_1)\sim reg.,
\label{eqn:q9}\\
&&E^{+,i}(z_1)S^j(z_2)=S^j(z_2)E^{+,i}(z_1)\sim reg.,
\label{eqn:q10}\\
&&E^{-,i}(z_1)S^j(z_2)=S^j(z_2)E^{-,i}(z_1)
\label{eqn:q11}\\
&&\sim reg.+\delta_{i,j}\times
~_{k+N}\partial_{z_2}\left(
\frac{1}{z_1-z_2}:e^{
\sum_{n \neq 0}\frac{a_n^i}{[(k+N)n]}q^{\frac{k+N}{2}|n|}z_2^{-n}
-\frac{1}{k+N}(q_a^i+p_a^i{\rm log}z_2)
}:\right).\nonumber
\end{eqnarray}
The symbol $\sim reg.$ means equality modulo regular function.
\end{prop}
The equalities (\ref{eqn:q1}), (\ref{eqn:q2}),
(\ref{eqn:q3}), (\ref{eqn:q4}), (\ref{eqn:q5}), (\ref{eqn:q6}),
(\ref{eqn:q7}), (\ref{eqn:q9}), (\ref{eqn:q10})
hold in "$\sim reg.$" sense.
The exceptional cases are (\ref{eqn:q8}) and (\ref{eqn:q11}),
which do not exist inside regular function.
Note that the elliptic theta function $[u]_{k+N}$ has already appeared 
in trigonometric symmetry 
$U_q(\widehat{sl_N})$.

\section{Free field realization of $U_{q,p}(\widehat{sl_N})$}

The purpose of this section is to give 
a free field realization for the elliptic algebra 
$U_{q,p}(\widehat{sl_N})$ for arbitrary level $k \neq 0, -N$.

\subsection{Drinfeld current}

Following \cite{CD},
let us introduce the auxiliary operators ${\cal B}_\pm^{* i,j}(z), 
{\cal B}_\pm^{i,j}(z)$,
$(1\leqq i<j \leqq N)$ by
\begin{eqnarray}
{\cal B}_\pm^{* i,j}(z)&=&\exp\left(\pm \sum_{n>0}\frac{1}{[r^*n]}b^{i,j}_{-n}
(q^{r^*-1}z)^{n}\right),
\\
{\cal B}_\pm^{i,j}(z)&=&\exp\left(\pm \sum_{n>0}\frac{1}{[rn]}b^{i,j}_n
(q^{-r^*+1}z)^{-n}\right).
\end{eqnarray}
Let us introduce the auxiliary operators ${\cal A}^{* i}(z), 
{\cal A}^i(z)$, 
$(1\leqq i \leqq N-1)$ by
\begin{eqnarray}
{\cal A}^{* i}(z)&=&
\exp\left(\sum_{n>0}\frac{1}{[r^*n]}a_{-n}^i (q^{r^*}z)^n\right),
\\
{\cal A}^{i}(z)&=&\exp\left(-\sum_{n>0}\frac{1}{[rn]}a_n^i 
(q^{-r^*}z)^{-n}\right).
\end{eqnarray}

\begin{df}~~We define the dressing operators 
${U}^{* i}(z), U^i(z),(1\leqq i \leqq N-1)$.
\begin{eqnarray}
U^{* i}(z)&=&\left(\prod_{j=1}^{i-1}
{\cal B}_+^{* j,i+1}(q^{2-j}z){\cal B}_-^{* j,i}(q^{1-j}z)\right)
\label{def:dress1}\\
&\times&{\cal B}_+^{* i,i+1}(q^{2-i}z){\cal B}_+^{* i,i+1}(q^{-i}z)
\left(\prod_{j=i+2}^N
{\cal B}_+^{* i,j}(q^{-j+1}z){\cal B}_-^{* i+1,j}(q^{-j+2}z)
\right){\cal A}^{* i}(q^{\frac{k-N}{2}}z),\nonumber
\\
{U}^i(z)&=&\left(\prod_{j=1}^{i-1}
{\cal B}_-^{j,i+1}(q^{-2+j}z){\cal B}_+^{j,i}(q^{-1+j}z)\right)
\label{def:dress2}
\\
&\times&{\cal B}_-^{i,i+1}(q^{-2+i}z){\cal B}_-^{i,i+1}(q^{i}z)
\left(\prod_{j=i+2}^N
{\cal B}_-^{i,j}(q^{j-1}z){\cal B}_+^{i+1,j}(q^{j-2}z)
\right){\cal A}^i(q^{\frac{-k+N}{2}}z).\nonumber
\end{eqnarray}
\end{df}
Formulae (\ref{def:dress1}) and (\ref{def:dress2}) are 
main result of this paper.
In appendix A we explain 
a systematic construction of
the dressing operators $U^{* i}(z), U^i(z)$.

\begin{df}~~
We define the elliptic deformation of Drinfeld current 
$e_i(z), f_i(z), \Psi_i^\pm(z), (1\leqq i \leqq N-1)$, by
\begin{eqnarray}
&&e_i(z)=U^{* i}(z)E^{+,i}(z),\\
&&f_i(z)=E^{-,i}(z)U^i(z),\\
&&\Psi_i^+(z)=U^{* i}(q^{\frac{k}{2}}z)\psi_i^+(z)U^i(q^{-\frac{k}{2}}z),\\
&&\Psi_i^-(z)=U^{* i}(q^{-\frac{k}{2}}z)\psi_i^-(z)U^i(q^{\frac{k}{2}}z).
\end{eqnarray}
\end{df}

~\\
{\bf Example}~~~~
Upon specialization $N=3$ we recover the dressing operator of 
$U_{q,p}(\widehat{sl_3})$
\cite{Kojima}. 
\begin{eqnarray}
U^{*1}(z)&=&{\cal B}_+^{* 1,2}(qz){\cal B}_+^{* 1,2}(q^{-1}z)
{\cal B}_+^{* 13}(q^{-2}z){\cal B}_-^{* 23}(q^{-1}z){\cal A}^{* 1}
(q^{\frac{k-3}{2}}z),\label{def:dsl3-1}\\
U^{*2}(z)&=&{\cal B}_+^{* 1,3}(qz)
{\cal B}_-^{* 1,2}(z){\cal B}_+^{* 2,3}(z)
{\cal B}_+^{* 2,3}(q^{-2}z)
{\cal A}^{* 2}(q^{\frac{k-3}{2}}z),\label{def:dsl3-2}\\
U^1(z)&=&{\cal B}_-^{1,2}(q^{-1}z){\cal B}_-^{12}(qz){\cal B}_-^{1,3}(q^2z)
{\cal B}_+^{2,3}(qz)
{\cal A}^1(q^{\frac{-k+3}{2}}z),\label{def:dsl3-3}
\\
U^2(z)&=&{\cal B}_-^{1,3}(q^{-1}z){\cal B}_+^{1,2}(z)
{\cal B}_-^{2,3}(z){\cal B}_-^{2,3}(q^2z)
{\cal A}^2(q^{\frac{-k+3}{2}}z),
\label{def:dsl3-4}
\end{eqnarray}
The notation of this paper is slightly different from those of \cite{Kojima}.
For example, ${\cal B}^{* 1,2}_\pm(z)={\cal B}^{* 1}_\pm(q^{r^*-1}z)$,
${\cal B}^{* 1,3}_\pm(z)={\cal B}^{* 2}_\pm(q^{r^*-1}z)$, 
${\cal B}^{* 2,3}_\pm(z)={\cal B}^{*,3}_\pm(q^{r^*-1}z)$.

\begin{prop}~~The bosonic operators $e_i(z), f_i(z), \Psi_i^\pm(z)$, 
$(1\leqq i \leqq N-1)$ satisfy
the following commutation relations.
\begin{eqnarray}
&&\Theta_{p^*}(q^{-A_{i,j}}z_1/z_2)
e_i(z_1)e_j(z_2)=q^{-A_{i,j}}\Theta_{p^*}(q^{A_{i,j}}z_1/z_2)
e_j(z_2)e_i(z_1),
\\
&&\Theta_{p}(q^{A_{i,j}}z_1/z_2)f_i(z_1)f_j(z_2)
=q^{A_{i,j}}
\Theta_{p}(q^{-A_{i,j}}z_1/z_2)
f_j(z_2)f_i(z_1),\\
&&\Theta_p(q^{A_{i,j}}z_1/z_2)\Theta_{p^*}(q^{-A_{i,j}}z_1/z_2)
\Psi_i^\pm(z_1)\Psi_j^\pm(z_2)\nonumber\\
&&=\Theta_p(q^{-A_{i,j}}z_1/z_2)
\Theta_{p^*}(q^{A_{i,j}}z_1/z_2)
\Psi_j^\pm(z_2)\Psi_i^\pm(z_1),\\
&&\Theta_p(pq^{A_{i,j}-k}z_1/z_2)
\Theta_{p^*}(p^*q^{-A_{i,j}+k}z_1/z_2)
\Psi_i^\pm(z_1)\Psi_j^\mp(z_2)\nonumber\\
&&=\Theta_p(pq^{-A_{i,j}-k}z_1/z_2)
\Theta_{p^*}(p^*q^{A_{i,j}+k}z_1/z_2)
\Psi_j^\mp(z_2)\Psi_i^\pm(z_1),
\\
&&
\Theta_{p^*}(q^{-A_{i,j}\pm \frac{k}{2}}z_1/z_2)
\Psi_i^\pm(z_1)e_j(z_2)=
\Theta_{p^*}(q^{A_{i,j}\pm \frac{k}{2}}z_1/z_2)
e_j(z_2)\Psi_i^\pm(z_1),\\
&&
\Theta_{p}(q^{A_{i,j}\mp \frac{k}{2}}z_1/z_2)
\Psi_i^\pm(z_1)f_j(z_2)=
\Theta_{p}(q^{-A_{i,j}\mp \frac{k}{2}}z_1/z_2)
f_j(z_2)\Psi_i^\pm(z_1),
\end{eqnarray}
\begin{eqnarray}
~[e_i(z_1),f_j(z_2)]=\frac{\delta_{i,j}}
{(q-q^{-1})z_1 z_2}\left(
\delta\left(q^{-k}\frac{z_1}{z_2}\right)
\Psi_i^+(q^{-k/2}z_1)-
\delta\left(q^{k}\frac{z_1}{z_2}\right)
\Psi_i^-(q^{-k/2}z_2)\right).
\end{eqnarray}
They satisfy Serre relation.
\begin{eqnarray}
&&(p^*q^2z_2/z_1:p^*)_\infty (p^*q^{-2}z_1/z_2;p^*)_\infty \nonumber\\
&&\times\left\{
(p^*q^{-1}z/z_1;p^*)_\infty 
(p^*q^{-1}z/z_2;p^*)_\infty 
(p^*qz_1/z;p^*)_\infty 
(p^*qz_1/z;p^*)_\infty 
e_i(z_1)e_i(z_2)e_j(z)\right.
\nonumber\\
&&-[2]
(p^*q^{-1}z/z_1;p^*)_\infty 
(p^*q^{-1}z_2/z;p^*)_\infty 
(p^*qz_1/z;p^*)_\infty 
(p^*qz/z_2;p^*)_\infty 
e_i(z_1)e_j(z)e_i(z_2)
\nonumber\\
&&\left.+(p^*q^{-1}z_1/z;p^*)_\infty 
(p^*q^{-1}z_2/z;p^*)_\infty 
(p^*qz/z_1;p^*)_\infty 
(p^*qz/z_2;p^*)_\infty 
e_j(z)e_i(z_1)e_i(z_2)\right\}
\nonumber\\
&&+(z_1\leftrightarrow z_2)=0,~~~~{\rm for}~A_{i,j}=-1,
\end{eqnarray}
\begin{eqnarray}
&&(pq^{-2}z_2/z_1:p)_\infty (pq^{2}z_1/z_2;p)_\infty \nonumber\\
&&\times\left\{
(pqz/z_1;p)_\infty 
(pqz/z_2;p)_\infty 
(pq^{-1}z_1/z;p)_\infty 
(pq^{-1}z_1/z;p)_\infty 
f_i(z_1)f_i(z_2)f_j(z)\right.
\nonumber\\
&&-[2]
(pqz/z_1;p)_\infty 
(pqz_2/z;p)_\infty 
(pq^{-1}z_1/z;p)_\infty 
(pq^{-1}z/z_2;p)_\infty 
f_i(z_1)f_j(z)f_i(z_2)
\nonumber\\
&&\left.+(pqz_1/z;p)_\infty 
(pqz_2/z;p)_\infty 
(pq^{-1}z/z_1;p)_\infty 
(pq^{-1}z/z_2;p)_\infty 
f_j(z)f_i(z_1)f_i(z_2)\right\}
\nonumber\\
&&+(z_1\leftrightarrow z_2)=0,~~~~{\rm for}~A_{i,j}=-1.
\end{eqnarray}
\end{prop}
Following \cite{Konno, JKOS2},
we introduce the Heisenberg algebra ${\cal H}$ generated by the following
$P_i,Q_i$, $(1\leqq i \leqq N-1)$.
\begin{eqnarray}
~[P_i,Q_j]=\frac{A_{i,j}}{2}.
\end{eqnarray}
\begin{df}~~Let us define the bosonic operators
$E_i(z), F_i(z), H_i^\pm(z)
\in U_q(\widehat{sl_N}){\otimes}{\cal H}$, $(1\leqq i \leqq N-1)$ by
\begin{eqnarray}
E_i(z)&=&e_1(z)e^{2Q_i}z^{-\frac{P_i-1}{r^*}},\\
F_j(z)&=&f_1(z)z^{\frac{h_i+P_i-1}{r}},
\\
H_i^\pm(z)&=&\Psi_i^\pm(z)e^{2Q_i}
q^{\mp h_i}
(q^{\pm (r-\frac{k}{2})}z)^{\frac{h_i+P_i-1}{r}-\frac{P_i-1}{r^*}}.
\end{eqnarray}
\end{df}

\begin{thm}~~The bosonic operators $E_i(z), F_i(z), H_i^\pm(z)$, 
$(1\leqq i \leqq N-1)$
satisfy the following commutation relations.
\begin{eqnarray}
&&
\left[u_1-u_2-\frac{A_{i,j}}{2}\right]_{r^*}
E_i(z_1)E_j(z_2)=
\left[u_1-u_2+\frac{A_{i,j}}{2}\right]_{r^*}E_j(z_2)E_i(z_1),
\label{eqn:e1}
\\
&&
\left[u_1-u_2+\frac{A_{i,j}}{2}\right]_{r}
F_i(z_1)F_j(z_2)=
\left[u_1-u_2-\frac{A_{i,j}}{2}\right]_{r}
F_j(z_2)F_i(z_1),
\label{eqn:e2}
\\
&&
\left[u_1-u_2+\frac{A_{i,j}}{2}\right]_r
\left[u_1-u_2-\frac{A_{i,j}}{2}\right]_{r^*}
H^\pm_i(z_1)H^\pm_j(z_2)\nonumber\\
&&=
\left[u_1-u_2-\frac{A_{i,j}}{2}\right]_r
\left[u_1-u_2+\frac{A_{i,j}}{2}\right]_{r^*}
H^\pm_j(z_2)H^\pm_i(z_1),\label{eqn:e3}
\\
&&
\left[u_1-u_2+\frac{A_{i,j}}{2}-\frac{k}{2}\right]_r
\left[u_1-u_2-\frac{A_{i,j}}{2}+\frac{k}{2}\right]_{r^*}
H^+_i(z_1)H^-_j(z_2)\nonumber\\
&&=
\left[u_1-u_2-\frac{A_{i,j}}{2}-\frac{k}{2}\right]_r
\left[u_1-u_2+\frac{A_{i,j}}{2}+\frac{k}{2}\right]_{r^*}
H^-_j(z_2)H^+_i(z_1),
\label{eqn:e4}\\
&&
\left[u_1-u_2\pm \frac{k}{4}-\frac{A_{i,j}}{2}\right]_{r^*}
H^\pm_i(z_1)E_j(z_2)=
\left[u_1-u_2\pm\frac{k}{4}+\frac{A_{i,j}}{2}\right]_{r^*}
E_j(z_2)H^\pm_i(z_1),\label{eqn:e5}
\\
&&
\left[u_1-u_2\mp \frac{k}{4}+\frac{A_{i,j}}{2}\right]_{r}
H^\pm_i(z_1)F_j(z_2)=
\left[u_1-u_2\mp\frac{k}{4}-\frac{A_{i,j}}{2}\right]_{r}
F_j(z_2)H^\pm_i(z_1),
\label{eqn:e6}
\end{eqnarray}
\begin{eqnarray}
~[E_i(z_1),F_j(z_2)]=\frac{\delta_{i,j}}{(q-q^{-1})z_1z_2}\left(
\delta\left(q^{-k}\frac{z_1}{z_2}\right)H_i^+(q^{-\frac{k}{2}}z_1)-
\delta\left(q^{k}\frac{z_1}{z_2}\right)H_i^-(q^{-\frac{k}{2}}z_2)\right).
\label{eqn:e7}
\end{eqnarray}
They satisfy Serre relation.
\begin{eqnarray}
&&z_1^{-\frac{1}{r^*}}
(p^*q^2z_2/z_1:p^*)_\infty (p^*q^{-2}z_1/z_2;p^*)_\infty \nonumber\\
&&\times\left\{
z_2^{\frac{1}{r^*}}z^{-\frac{1}{r^*}}
(p^*q^{-1}z/z_1;p^*)_\infty 
(p^*q^{-1}z/z_2;p^*)_\infty 
(p^*qz_1/z;p^*)_\infty 
(p^*qz_1/z;p^*)_\infty 
E_i(z_1)E_i(z_2)E_j(z)\right.
\nonumber\\
&&-[2]
(p^*q^{-1}z/z_1;p^*)_\infty 
(p^*q^{-1}z_2/z;p^*)_\infty 
(p^*qz_1/z;p^*)_\infty 
(p^*qz/z_2;p^*)_\infty 
E_i(z_1)E_j(z)E_i(z_2)
\nonumber\\
&&\left.+
z^{\frac{1}{r^*}}z_1^{-\frac{1}{r^*}}(p^*q^{-1}z_1/z;p^*)_\infty 
(p^*q^{-1}z_2/z;p^*)_\infty 
(p^*qz/z_1;p^*)_\infty 
(p^*qz/z_2;p^*)_\infty 
E_j(z)E_i(z_1)E_i(z_2)\right\}
\nonumber\\
&&+(z_1\leftrightarrow z_2)=0,~~~~{\rm for}~A_{i,j}=-1,
\label{eqn:e8}
\\
&&z_1^{\frac{1}{r}}
(pq^{-2}z_2/z_1:p)_\infty (pq^{2}z_1/z_2;p)_\infty \nonumber\\
&&\times\left\{
z^{\frac{1}{r}}z_2^{-\frac{1}{r}}
(pqz/z_1;p)_\infty 
(pqz/z_2;p)_\infty 
(pq^{-1}z_1/z;p)_\infty 
(pq^{-1}z_1/z;p)_\infty 
F_i(z_1)F_i(z_2)F_j(z)\right.
\nonumber\\
&&-[2]
(pqz/z_1;p)_\infty 
(pqz_2/z;p)_\infty 
(pq^{-1}z_1/z;p)_\infty 
(pq^{-1}z/z_2;p)_\infty 
F_i(z_1)F_j(z)F_i(z_2)
\nonumber\\
&&\left.+
z_1^{\frac{1}{r}}z^{-\frac{1}{r}}
(pqz_1/z;p)_\infty 
(pqz_2/z;p)_\infty 
(pq^{-1}z/z_1;p)_\infty 
(pq^{-1}z/z_2;p)_\infty 
F_j(z)F_i(z_1)F_i(z_2)\right\}
\nonumber\\
&&+(z_1\leftrightarrow z_2)=0,~~~~{\rm for}~A_{i,j}=-1.
\label{eqn:e9}
\end{eqnarray}
\end{thm}
Now we have constructed level $k$ 
free field realization of Drinfeld current 
$E_i(z), F_i(z), H_i^\pm(z)$
for the elliptic algebra
$U_{q,p}(\widehat{sl_N})$ \cite{JKOS2, KK}.

\subsection{Screening current}

In this section we study the screening current for $U_{q,p}(\widehat{sl_N})$.
In the paper \cite{Konno} it was recognized that
the screening current of $U_{q,p}(\widehat{sl_2})$ was 
exactly the same as those of
$U_q(\widehat{sl_2})$.
Hence we select the same definition of screening current of 
$U_q(\widehat{sl_N})$ \cite{AOS2} as the screening current of 
$U_{q,p}(\widehat{sl_N})$.
\begin{eqnarray}
S_i(z)&=&\frac{-1}{(q-q^{-1})z}:e^{-a^i(z)}:\nonumber\\
&\times& \sum_{j=i+1}^N
:e^{\gamma^{i+1,j}(q^{N-j}z)}(
e^{-\beta_4^{i,j}(q^{N-j}z)}
-
e^{-\beta_3^{i,j}(q^{N-j}z)}
)e^{\sum_{l=j+1}^N
(b_-^{i+1,l}(q^{N-l+1}z)-b_-^{i,l}(q^{N-l}z))
}:\nonumber
\end{eqnarray}

\begin{prop}~~
The bosonic operator $S_i(z), E_i(z), F_i(z)$, $(1\leqq i \leqq N-1)$ satisfy
the following commutation relations.
\begin{eqnarray}
&&\left[u_1-u_2-\frac{A_{i,j}}{2}\right]_{k+N} 
S_i(z_1)S_j(z_2)=
\left[u_1-u_2+\frac{A_{i,j}}{2}\right]_{k+N} S_j(z_2)S_i(z_1)\sim reg.,
\label{eqn:e10}\\
&&
\left[u_1-u_2-\frac{A_{i,j}}{2}\right]_{r-k}
E_i(z_1)E_j(z_2)=
\left[u_1-u_2+\frac{A_{i,j}}{2}\right]_{r-k}E_j(z_2)E_i(z_1)\sim reg.,
\label{eqn:e11}\\
&&
\left[u_1-u_2+\frac{A_{i,j}}{2}\right]_{r}
F_i(z_1)F_j(z_2)=
\left[u_1-u_2-\frac{A_{i,j}}{2}\right]_{r}
F_j(z_2)F_i(z_1) \sim reg..
\label{eqn:e12}
\end{eqnarray}

\begin{eqnarray}
&&E_i(z_1)S_j(z_2)=S_j(z_2)E_i(z_1)\sim reg.,
\label{eqn:e13}
\end{eqnarray}
\begin{eqnarray}
&&F_i(z_1)S_j(z_2)=S_j(z_2)F_i(z_1)
\label{eqn:e14}\\
&&\sim reg.+\delta_{i,j}\times ~_{k+N}\partial_{z_2}\left(
\frac{1}{z_1-z_2}:e^{
\sum_{n \neq 0}\frac{a_n^i}{[(k+N)n]}q^{\frac{k+N}{2}|n|}z_2^{-n}
-\frac{1}{k+N}(q_a^i+p_a^i{\rm log}z_2)
}U^i(z_2)z_2^{\frac{h_i+P_i-1}{r}}:\right).\nonumber
\end{eqnarray}
The symbol $\sim reg.$ means eqality modulo regular function.
\end{prop}
The equalities (\ref{eqn:e1}), (\ref{eqn:e2}),
(\ref{eqn:e3}), (\ref{eqn:e4}), (\ref{eqn:e5}), (\ref{eqn:e6}),
(\ref{eqn:e8}), (\ref{eqn:e9}), (\ref{eqn:e10}), (\ref{eqn:e11})
hold in "$\sim reg.$" sense.
The exceptional cases are (\ref{eqn:e7}) and (\ref{eqn:e14}),
which do not exist inside regular function.
It seems to be possible to construct
three kind of infinitly many commutative operators,
which are baesd on
the commutation relations (\ref{eqn:e10}), (\ref{eqn:e11}), 
(\ref{eqn:e12}).
See references \cite{KS1, KS2, KS3}.

\section*{Acknowledgement}~The author would like to thank
Dr.S.Nagoya for his pointing out reference of Wakimoto realization.
The author would like to thank the organizing committee
of International Conference on Nonlinear Evolution Equations and Wave
Phenomena for giving me an opportunity to give a talk.
This work is partly supported by the Grant-in Aid
for Young Scientist {\bf B}(18740092) from Japan Society for
the Promotion of Science.

~\\

\begin{appendix}

\section{Construction of dressing operators $U^i(z), U^{* i}(z)$}

In this appendix 
we explain a systematic way 
to find the dressing operators $U^i(z)$and $U^{* i}(z)$
associated with the elliptic algebra $U_{q,p}(\widehat{sl_N})$.
Wakimoto realization is not symmetric with Cartan subalgebra.
In other words,
Wakimoto realization of Drinfeld current $E^{+,i}(z)$ 
is very different from those of 
$E^{-,i}(z)$.
The realization of $E^{+,i}(z)$ is simpler than those of $E^{-,i}(z)$.
Hence it is better to
consider the dressing operator $U^{* i}(z)$ associated with
$E^{+,i}(z)$ at first.
We construct the dressing operator $U^{*i}(z)$ by products of
the basic operator ${\cal B}^{* i,j}_\pm(z)$.
The commutation relation
between $E^{\pm i}_j(z)$ and ${\cal B}^{* k,l}_\pm(z)$ are complicated.
Hence we prepare auxiliary operators $\widetilde{\cal B}^{* i,j}_+(z)$
which commute with at most every Drinfeld current $E_l^{+,k}(z)$.
\begin{eqnarray}
[\widetilde{\cal B}^{* j,i+1}_+(z_1),E_j^{+,i}(z_2)]\neq 0,~~
[\widetilde{\cal B}^{* j,i+1}_+(z_1),E_l^{+,k}(z_2)]=0,~{\rm for}~(k,l)\neq (i,j).
\end{eqnarray}
For example, the explicit formulae of $\widetilde{\cal B}^{* i,j}_+(z)$
for $U_{q,p}(\widehat{sl_4})$ are given as followings.
\begin{eqnarray}
\widetilde{\cal B}^{* 1,2}_+(z)&=&
{\cal B}_+^{* 1,2}(z)
{\cal B}_+^{* 1,3}(q^{-1}z)
{\cal B}_+^{* 1,4}(q^{-2}z),\nonumber
\\
\widetilde{\cal B}^{* 1,3}_+(z)&=&
{\cal B}_+^{* 1,3}(z)
{\cal B}_+^{* 1,4}(q^{-1}z)
{\cal B}_+^{* 2,3}(q^{-1}z)
{\cal B}_+^{* 2,4}(q^{-2}z)
{\cal B}_-^{* 2,3}(qz)
{\cal B}_-^{* 2,4}(z),\nonumber
\\
\widetilde{\cal B}^{* 2,3}_+(z)&=&
{\cal B}_+^{* 2,3}(z)
{\cal B}_+^{* 2,4}(q^{-1}z),\nonumber
\\
\widetilde{\cal B}^{* 1,4}_+(z)&=&
{\cal B}_+^{* 1,4}(z)
{\cal B}_+^{* 2,4}(q^{-1}z)
{\cal B}_+^{* 3,4}(q^{-2}z)
{\cal B}_-^{* 2,4}(qz)
{\cal B}_-^{* 3,4}(z),\nonumber
\\
\widetilde{\cal B}^{* 2,4}_+(z)&=&
{\cal B}_+^{* 2,4}(z)
{\cal B}_+^{* 3,4}(q^{-1}z)
{\cal B}_-^{* 3,4}(qz),\nonumber\\
\widetilde{\cal B}^{* 3,4}_+(z)&=&
{\cal B}_+^{* 3,4}(z).\nonumber
\end{eqnarray}
The remaining non-commutative commutation relation is given by
\begin{eqnarray}
E_j^{+ i-1}(z_1)
{\cal B}^{* i,j}_+(q^{j-1}z_1)=
\frac{
(p^*q^{-1}z_2/z_1;p^*)_\infty }{
(p^*qz_2/z_1;p^*)_\infty}
{\cal B}^{* i,j}_+(q^{j-1}z_1)
E_j^{+ j-1}(z_1).\nonumber
\end{eqnarray}
For simplicity, 
we demonstrate this construction in $U_{q,p}(\widehat{sl_4})$ case.
The commutation relation between $:e^{\beta_2^{i,j}(z_1)}:$ and 
${\cal B}^{* i,j}_\pm(z_2)$ is exactly
the same as those between
$:e^{\beta_1^{i,j}(z_1)}:$ and 
${\cal B}^{* i,j}_\pm(z_2)$.
Hence, in what follows, we can regard
\begin{eqnarray}
&&E^{+,1}_1(z) \sim :e^{\beta_1^{12}(z)}:,\nonumber\\
&&E^{+,2}_1(z) \sim :e^{\gamma^{12}(z)+\beta_1^{13}(z)}:,\nonumber\\
&&E^{+,2}_2(z) \sim :e^{\beta_1^{23}(qz)+b_+^{13}(z)-b_+^{12}(qz)}:,\nonumber\\
&&E^{+,3}_1(z) \sim :e^{\gamma^{13}(z)+\beta_1^{14}(z)}:,\nonumber\\
&&E^{+,3}_2(z) \sim :e^{\gamma^{23}(qz)+\beta_1^{24}(qz)
+b_+^{14}(z)-b_+^{13}(qz)}:,\nonumber\\
&&E^{+,3}_3(z) \sim :e^{\beta_1^{34}(q^2z)+b_+^{14}(z)-b_+^{13}(qz)
+b_+^{24}(qz)-b_+^{23}(q^2z)}:.\nonumber
\end{eqnarray}
There exists lexicographical ordering structure for index $(i,j)$ of
$b_m^{i,j}$ inside $E^{+,i}(z)$.
Hence we assume the formulae of 
$\widetilde{\cal B}^{* i,j}_+(z)$ as following.
\begin{eqnarray}
\widetilde{\cal B}^{* i,j}_+(z)={\cal B}^{* i,j}_+(z) \times 
\prod_{(k,l)
\atop{(i,j)<(k,l)}}{\cal B}_+^{* k,l}(q^{m_{k,l}^+}z)^{\epsilon_{k,l}^+}
{\cal B}_-^{* k,l}(q^{m_{k,l}^-}z)^{\epsilon_{k,l}^-}.\label{eqn:app2}
\end{eqnarray}
Here $m_{k,l}^\pm \in {\mathbb Z}$ and $\epsilon_{k,l}^\pm \in {\mathbb N}$.
Here $(i,j)<(k,l)$ means 
the lexicographical ordering.i.e. $(1,2)<(1,3)<(1,4)<(2,3)<(2,4)<(3,4)$.
\\
$\bullet$~Let's determine 
$\widetilde{\cal B}^{* 12}_+(z)={\cal B}^{* 12}_+(z)\times 
\cdots $. 
In order to satisfy the commutativity
$[{\cal B}^{* 12}_+(z_1),E^{+,2}_1(z_2)]=0$,
the auxiliary operator should be
$\widetilde{\cal B}^{* 12}_+(z)={\cal B}_+^{* 12}(z){\cal B}^{* 13}_+(q^{-1}z) \cdots $.
Upon this assumption, the commutativity
$[{\cal B}^{* 12}_+(z_1),E^{+,2}_2(z_2)]=0$ holds automatically.
In order to satisfy the commutativity
$[{\cal B}^{* 12}_+(z_1),E^{+,3}_1(z_2)]=0$,
the auxiliary operator should be
$\widetilde{\cal B}^{* 12}_+(z)={\cal B}_+^{* 12}(z){\cal B}^{* 13}_+(q^{-1}z)
{\cal B}_+^{* 14}(q^{-2}z) \cdots $.
Upon this assumption, the commutation relation
$[{\cal B}^{* 12}_+(z_1),E^{+,3}_2(z_2)]=0$ and
$[{\cal B}^{* 12}_+(z_1),E^{+,3}_3(z_2)]=0$
hold automatically.
Hence we conclude
$\widetilde{\cal B}^{* 12}_+(z)={\cal B}_+^{* 12}(z){\cal B}^{* 13}_+(q^{-1}z)
{\cal B}_+^{* 14}(q^{-2}z)$.
The auxiliary operator
$\widetilde{\cal B}^{* i,i+1}_+(z)$ 
is determined as the same manner.
\\
$\bullet$~Let's determine $\widetilde{\cal B}^{* 13}_+(z)={\cal B}^{* 13}_+(z)
\times \cdots $.
Because of the assumption (\ref{eqn:app2}),
commutativity
$[\widetilde{\cal B}^{* 13}_+(z_1), E^{+,1}_1(z_2)]=0$ holds.
In order to satisfy the commutativity
$[\widetilde{\cal B}^{* 13}_+(z_1),E^{+,2}_2(z_2)]=0$,
the auxiliary operator should be
$\widetilde{\cal B}^{* 13}_+(z)=
{\cal B}_+^{* 13}(z){\cal B}^{* 23}_+(q^{-1}z){\cal B}^{* 23}_-(qz) \cdots $.
In order to satisfy the commutativity
$[\widetilde{\cal B}^{* 13}_+(z_1),E^{+,3}_1(z_2)]=0$,
the dressing operator should be
$\widetilde{\cal B}^{* 13}_+(z)=
{\cal B}_+^{* 13}(z){\cal B}^{* 23}_+(q^{-1}z){\cal B}^{* 23}_-(qz)
{\cal B}_+^{* 14}(q^{-1}z) \cdots $.
In order to satisfy the commutativity
$[\widetilde{\cal B}^{* 13}_+(z_1),E^{+,3}_2(z_2)]=0$,
the auxiliary operator should be
$\widetilde{\cal B}^{* 13}_+(z)=
{\cal B}_+^{* 13}(z){\cal B}^{* 23}_+(q^{-1}z){\cal B}^{* 23}_-(qz)
{\cal B}_+^{* 14}(q^{-1}z) 
{\cal B}^{* 24}_+(q^{-2}z){\cal B}^{* 24}_-(z)\times 
\cdots $.
Upon these assumption,
the commutativity
$[\widetilde{\cal B}^{* 13}(z_1),E^{+,3}_3(z_2)]=0$ holds, automatically.
Hence we conclude $
\widetilde{\cal B}^{* 13}_+(z)=
{\cal B}_+^{* 13}(z){\cal B}^{* 23}_+(q^{-1}z){\cal B}^{* 23}_-(qz)
{\cal B}_+^{* 14}(q^{-1}z) 
{\cal B}^{* 24}_+(q^{-2}z){\cal B}^{* 24}_-(z)$.
The auxiliary operator $\widetilde{\cal B}^{* i, i+2}_+(z)$
is determined as the same manner.\\
$\bullet$~Let's determine $\widetilde{\cal B}^{* 14}_+(z)={\cal B}^{* 14}_+(z)
\times \cdots $.
Because of the assumption (\ref{eqn:app2}),
the commutation relations $[\widetilde{B}^{* 14}_+(z_1),E^{+ 1}(z_2)]=
[\widetilde{B}^{* 14}_+(z_1),E^{+ 2}(z_2)]=0$ hold.
In order to satisfy the commutativity
$[\widetilde{\cal B}^{* 14}_+(z_1),E^{+,3}_2(z_2)]=0$,
the auxiliary operator should be
$\widetilde{\cal B}^{* 14}_+(z)=
{\cal B}_+^{* 14}(z){\cal B}^{* 24}_+(q^{-1}z){\cal B}^{* 24}_-(qz)
\cdots $.
In order to satisfy the commutativity
$[\widetilde{\cal B}^{* 14}_+(z_1),E^{+,3}_3(z_2)]=0$,
the dressing operator should be
$\widetilde{\cal B}^{* 14}_+(z)=
{\cal B}_+^{* 14}(z)
{\cal B}^{* 24}_+(q^{-1}z){\cal B}^{* 24}_-(qz)
{\cal B}^{* 34}_+(q^{-2}z){\cal B}^{* 34}_-(z)$.
The dressing operator $\widetilde{\cal B}^{* i, i+3}_+(z)$
is determined as the same manner.

We have determined the auxiliary operators 
$\widetilde{\cal B}^{i,j}_+(z)$ for $U_{q,p}(\widehat{sl_4})$.

As you have seen the above,
the lexicographical ordering structure 
inside $E^{+,i}(z)$ plays an important role
in construction of the auxiliary operator $\widetilde{\cal B}^{* ij}_+(z)$.
As the same manner as the above, we have the explicit formulae
of the auxiliary operator ${\cal B}^{* i,j}_+(z)$ for 
the elliptic algebra $U_{q,p}(\widehat{sl_N})$ as following.
\begin{eqnarray}
\widetilde{\cal B}^{* i,j}_+(z)=\prod_{s=i}^{j-1}\prod_{t=j}^N 
{\cal B}_+^{* s,t}(q^{i+j-s-t}z)\prod_{s=i+1}^{j-1}\prod_{t=j}^N 
{\cal B}_-^{* s,t}(q^{i+j+2-s-t}z),
\end{eqnarray}
We have the commutation relation.
\begin{eqnarray}
&&\Theta_{p^*}(q^{-1}z_1/z_2)E_j^{+,i}(z_1)\widetilde{\cal B}^{* i+1,j}_+(z_1)
E_j^{+,i}(z_2)\widetilde{\cal B}^{* i+1,j}(z_2)\nonumber\\
&&=q^{-1}
\Theta_{p^*}(q z_1/z_2)E_j^{+,i}(z_2)\widetilde{\cal B}^{* i+1,j}_+(z_2)
E_j^{+,i}(z_1)\widetilde{\cal B}^{* i+1,j}(z_1).\label{eqn:app3}
\end{eqnarray}
Let us set the auxiliary operator
$\widetilde{\cal B}^{* i,j}_-(z)=\prod_{s=i}^{j-1}\prod_{t=j}^N 
{\cal B}_-^{* s,t}(q^{i+j-s-t}z)\prod_{s=i+1}^{j-1}\prod_{t=j}^N 
{\cal B}_+^{* s,t}(q^{i+j+2-s-t}z)$.
Considering about
the equation (\ref{eqn:app3})
and the structure of Cartan matrix of the classical $sl_N$,
we set the dressing operator
\begin{eqnarray}
\widetilde{U}^{* i}(z)=\prod_{j=1}^i 
\widetilde{B}_+^{* j,i+1}(q^jz)
\widetilde{B}_+^{* j,i+1}(q^{j-2}z)
\prod_{j=1}^{i-1}
\widetilde{B}_-^{* j,i}(q^{j-1}z)
\prod_{j=1}^{i+1}
\widetilde{B}_-^{* j,i+1}(q^{j-1}z).\label{eqn:app4}
\end{eqnarray}
Let us set $\widetilde{e}_i(z)
=\widetilde{U}^{* i}(z)E^{+,i}(z)$, $(1\leqq i \leqq N-1)$.
We have the commutation relations
\begin{eqnarray}
\Theta_{p^*}(q^{-A_{i,j}}z_1/z_2)
\widetilde{e}_i(z_1)\widetilde{e}_j(z_2)=
q^{-A_{i,j}}\Theta_{p^*}(q^{A_{i,j}}z_1/z_2)
\widetilde{e}_j(z_2)\widetilde{e}_i(z_1).\nonumber
\end{eqnarray}
Clearing up overlap, we have 
\begin{eqnarray}
\widetilde{U}^{* i}(z)&=&
\left(\prod_{j=1}^{i-1}
{\cal B}_+^{* j,i+1}(q^{2-j}z){\cal B}_-^{* j,i}(q^{1-j}z)\right)
\nonumber
\\
&\times&{\cal B}_+^{* i,i+1}(q^{2-i}z){\cal B}_+^{* i,i+1}(q^{-i}z)
\left(\prod_{j=i+2}^N
{\cal B}_+^{* i,j}(q^{-j+1}z){\cal B}_-^{* i+1,j}(q^{-j+2}z)\right).\nonumber
\end{eqnarray}
Next we consider the dressing operator $U^{i}(z)$.
The structure of $E^{-,i}(z)$ is more complicated than
those of $E^{+,i}(z)$.
It is difficult to use lexicorgaphical ordering structure for $E^{-,i}(z)$.
Now let's go back to the explicit formulae
of the dressing operator for $U_{q,p}(\widehat{sl_3})$,
(\ref{def:dsl3-1}), (\ref{def:dsl3-2}), (\ref{def:dsl3-3}), 
(\ref{def:dsl3-4}) \cite{Kojima}.
There exists "duality" relation ${\cal B}_\pm^{* i,j}(q^sz)\leftrightarrow
{\cal B}_\mp^{i,j}(q^{-s}z)$ between the dressing operators
$U^{* i}(z)$ and $U^i(z)$ for $U_{q,p}(\widehat{sl_3})$.
Hence we set
\begin{eqnarray}
\widetilde{U}^{i}(z)&=&
\left(\prod_{j=1}^{i-1}
{\cal B}_-^{j,i+1}(q^{-2+j}z){\cal B}_+^{j,i}(q^{-1+j}z)\right)
\nonumber
\\
&\times&{\cal B}_-^{i,i+1}(q^{-2+i}z)
{\cal B}_-^{i,i+1}(q^{i}z)
\left(\prod_{j=i+2}^N
{\cal B}_-^{i,j}(q^{j-1}z){\cal B}_+^{i+1,j}(q^{j-2}z)\right).\nonumber
\end{eqnarray}
Let us set $e_i(z), f_i(z), \Psi_i^\pm(z)$, 
$(1\leqq i \leqq N-1)$ by
\begin{eqnarray}
&&e_i(z)=U^{* i}(z)E^{+,i}(z),~f_i(z)=E^{-,i}(z)U^i(z),\nonumber\\
&&\Psi_i^+(z)=U^{*i}(q^{\frac{k}{2}}z)\psi^+_i(z)U^i(q^{-\frac{k}{2}}z),~
\Psi_i^-(z)=U^{*i}(q^{-\frac{k}{2}}z)
\psi^-_i(z)U^i(q^{\frac{k}{2}}z).\nonumber
\nonumber
\end{eqnarray}
where we have set
\begin{eqnarray}
U^{* i}(z)=\widetilde{U}^{* i}(z){\cal A}^{* i}(q^{s_i}z),~~~
U^{i}(z)=\widetilde{U}^{i}(z){\cal A}^{i}(q^{-s_i}z),~~
(s_i \in {\mathbb R}).\nonumber
\end{eqnarray}
By necessary condition on commutation relations
we get the parameters $s_i=\frac{k-N}{2}$.
Now we have gotten conjecturous formulae of the dressing operators 
$U^{* i}(z)$ and $U^i(z)$.
Using appendix B, we can show every commutation relations
of $e_i(z), f_i(z), \Psi_i^\pm(z)$, by direct calculation.
It seems that the method explained above
can be applied to
the elliptic algebra $U_{q,p}({\goth g})$ for arbitrary ${\goth g}$.

\section{Normal ordering}

In this appendix we summarize the normal ordering of the basic operators.
\begin{eqnarray}
~:e^{\beta_1^{i,j}(z_1)}:
{\cal B}_+^{* i,j}(z_2)&=&::\frac{(q^{2r^*-1}z_2/z_1;q^{2r^*})_\infty}{
(q^{2r^*+1}z_2/z_1;q^{2r^*})_\infty},\nonumber\\
~:e^{\beta_2^{i,j}(z_1)}:
{\cal B}_+^{* i,j}(z_2)&=&::\frac{(q^{2r^*-1}z_2/z_1;q^{2r^*})_\infty}{
(q^{2r^*+1}z_2/z_1;q^{2r^*})_\infty},\nonumber\\
~:e^{\beta_3^{i,j}(z_1)}:
{\cal B}_+^{* i,j}(z_2)&=&::\frac{(q^{2r^*-1}z_2/z_1;q^{2r^*})_\infty}{
(q^{2r^*-3}z_2/z_1;q^{2r^*})_\infty},\nonumber\\
~:e^{\beta_4^{i,j}(z_1)}:
{\cal B}_+^{* i,j}(z_2)&=&::\frac{(q^{2r^*-1}z_2/z_1;q^{2r^*})_\infty}{
(q^{2r^*-3}z_2/z_1;q^{2r^*})_\infty},\nonumber\\
:e^{\gamma^{i,j}(z_1)}:
{\cal B}_+^{* i,j}(z_2)&=&::\frac{(q^{2r^*}z_2/z_1;q^{2r^*})_\infty}{
(q^{2r^*-2}z_2/z_1;q^{2r^*})_\infty},\nonumber
\\
:e^{b_+^{i,j}(z_1)}:
{\cal B}_+^{* i,j}(z_2)&=&::\frac{(q^{2r^*-1}z_2/z_1;q^{2r^*})_\infty^2}{
(q^{2r^*+1}z_2/z_1;q^{2r^*})_\infty 
(q^{2r^*-3}z_2/z_1;q^{2r^*})_\infty},\nonumber
\\
{\cal B}_-^{i,j}(z_1)
:e^{\beta_1^{i,j}(z_2)}:&=&::\frac{(q^{2r-k-1}z_2/z_1;q^{2r})_\infty}{
(q^{2r-k+1}z_2/z_1;q^{2r})_\infty},\nonumber\\
{\cal B}_-^{i,j}(z_1)
:e^{\beta_2^{i,j}(z_2)}:&=&::\frac{(q^{2r-k-1}z_2/z_1;q^{2r})_\infty}{
(q^{2r-k+1}z_2/z_1;q^{2r})_\infty},\nonumber\\
{\cal B}_-^{i,j}(z_1)
:e^{\beta_3^{i,j}(z_2)}:&=&
::\frac{(q^{2r-k-1}z_2/z_1;q^{2r})_\infty}{
(q^{2r-k-3}z_2/z_1;q^{2r})_\infty},\nonumber\\
{\cal B}_-^{i,j}(z_1)
:e^{\beta_4^{i,j}(z_2)}:&=&::\frac{(q^{2r-k-1}z_2/z_1;q^{2r})_\infty}{
(q^{2r-k-3}z_2/z_1;q^{2r})_\infty},\nonumber\\
{\cal B}_-^{i,j}(z_1)
:e^{\gamma^{i,j}(z_2)}:&=&::\frac{(q^{2r-k}z_2/z_1;q^{2r})_\infty}{
(q^{2r-k-2}z_2/z_1;q^{2r})_\infty},\nonumber\\
{\cal B}_-^{i,j}(z_1)
:e^{b_-^{i,j}(z_2)}:&=&::\frac{(q^{2r-k-1}z_2/z_1;q^{2r})_\infty^2}{
(q^{2r-k+1}z_2/z_1;q^{2r})_\infty (q^{2r-k-3}z_2/z_1;q^{2r})_\infty},\nonumber
\\
{\cal A}^i(z_1):e^{a_-^j(z_2)}:&=&::
\frac{(q^{2r+N+A_{i,j}}z_2/z_1;q^{2r})_\infty 
(q^{2r-2k-N-A_{i,j}}z_2/z_1;q^{2r})_\infty}
{(q^{2r+N-A_{i,j}}z_2/z_1;q^{2r})_\infty 
(q^{2r-2k-N+A_{i,j}}z_2/z_1;q^{2r})_\infty},\nonumber\\
:e^{a_+^i(z_1)}:{\cal A}^{* j}(z_2)
&=&::
\frac{(q^{2r^*+N+A_{i,j}}z_2/z_1;q^{2r^*})_\infty 
(q^{2r^*-2k-N-A_{i,j}}z_2/z_1;q^{2r^*})_\infty}
{(q^{2r^*+N-A_{i,j}}z_2/z_1;q^{2r^*})_\infty 
(q^{2r^*-2k-N+A_{i,j}}z_2/z_1;q^{2r^*})_\infty},\nonumber\\
{\cal B}_-^{i,j}(z_1){\cal B}_{+}^{* i,j}(z_2)&=&:
:\frac{(q^k z_2/z_1;q^{2k},p^*)_\infty^2}{
(q^{k+2}z_2/z_1;q^{2k},q^{2r^*})_\infty
(q^{k-2}z_2/z_1;q^{2k},q^{2r^*})_\infty
}\nonumber\\
&\times&\frac{
(q^{k+2}z_2/z_1;q^{2k},q^{2r})_\infty
(q^{k-2}z_2/z_1;q^{2k},q^{2r})_\infty}
{(q^k z_2/z_1;q^{2k},q^{2r})_\infty^2},
\nonumber\\
{\cal A}^i(z_1){\cal A}^{*j}(z_2)&=&::
\frac{(q^{2k+N+A_{i,j}}z_2/z_1;q^{2k},q^{2r^*})_\infty
(q^{-N-A_{i,j}}z_2/z_1;q^{2k},q^{2r^*})_\infty
}{
(q^{2k+N-A_{i,j}}z_2/z_1;q^{2k},q^{2r^*})_\infty
(q^{-N+A_{i,j}}z_2/z_1;q^{2k},q^{2r^*})_\infty
}\nonumber\\
&\times&
\frac{(q^{2k+N-A_{i,j}}z_2/z_1;q^{2k},q^{2r})_\infty
(q^{-N+A_{i,j}}z_2/z_1;q^{2k},q^{2r})_\infty
}{
(q^{2k+N+A_{i,j}}z_2/z_1;q^{2k},q^{2r})_\infty
(q^{-N-A_{i,j})}z_2/z_1;q^{2k},q^{2r})_\infty}.\nonumber
\end{eqnarray}
Here we have used the notation 
$$(z;p_1,p_2)_\infty=\prod_{n_1,n_2=0}^\infty
(1-p_1^{n_1}p_2^{n_2}z).$$

\end{appendix}

\end{document}